\documentstyle[epsf,a4,12pt]{article}
\begin{document}

\begin{titlepage}
\rightline {Si-97-09 \  \  \  \   }

\vspace*{2.truecm}

\centerline{\Large \bf  Critical Relaxation and Critical Exponents 
\footnote
{Work supported in part by the Deutsche Forschungsgemeinschaft;
 DFG~Schu 95/9-1}}
\vskip 0.6truecm

\vskip 2.0truecm
\centerline{\bf H.J. Luo\footnote
{\noindent On leave of absence from Sichua Union University,
 Chengdu, P.R. China} and B.  Zheng}
\vskip 0.2truecm

\vskip 0.2truecm
\centerline{Universit\"at -- GH Siegen, D -- 57068 Siegen, Germany}

\vskip 2.5truecm

\abstract{Dynamic relaxation of the XY model and fully frustrated
XY model quenched from
an initial ordered state to the critical temperature or below
is investigated with Monte Carlo methods. 
Universal power law scaling behaviour is observed.
The dynamic critical exponent $z$ and the static exponent
$\eta$ are extracted from the time-dependent
Binder cumulant and magnetization. The results are
 competitive to those measured with traditional methods. }

\vspace{0.5cm}


\end{titlepage}

Much progress has recently been achieved
in critical dynamics.
For a critical relaxation process starting from
a disordered state,
it was recently argued by Janssen, Schaub and Schmittmann
\cite {jan89}
with renormalization group methods that
there exists
universal scaling behaviour even {\it at macroscopic early
times},
 which set in right after a
microscopic time scale $t_{mic}$.
Important is that
extra critical exponents should be introduced
to describe the dependence of the scaling behaviour on the initial
conditions \cite {jan89,hus89},
 or to characterize the scaling behaviour of
special dynamic observables \cite {maj96,sch97,oer97}.
The critical exponent $\theta$ governing the
initial increase of the magnetization and the exponent
$\theta_1$
characterizing the power law decay of the 
global persistence probability have numerically been determined
to a satisfactory level for the Ising and the Potts model
\cite {oka97,gra95,li94}.

The short-time dynamic scaling 
is found to be quite general,
for example, for the dynamics beyond model A \cite {oer93},
 at tricritical point \cite {oer94}
and in surface critical
phenomena \cite {rit95} as well as with even more arbitrary
initial conditions \cite {li95,sch96,zhe96}. 
The investigation of the universal behaviour of
the short-time dynamics not only enlarges the fundamental
knowledge
on critical phenomena but also, more interestingly,
provides possible new ways to determine all the
dynamic exponents as well as the static exponents.
Especially, it has been demonstrated with the 
two-dimensional Ising model and Potts model
that the extraction of the exponents from
the power law behaviour of the observables
at the beginning of the time evolution is
rather efficient \cite {sch95,gra95,sch96,oka97a}.
 Since our measurements do not
enter the long-time regime of the dynamic
evolution, the method may be free of critical slowing down.
Therefore, it is interesting and important to
investigate its application to complex systems
where critical slowing down is more severe.

The two-dimensional XY model and fully frustrated
XY model (FFXY) are two typical examples where
numerical simulations are hampered
by the severe critical slowing down.
The dynamic relaxation
of these two models starting from {\it a disordered state} with 
a small magnetization
has been investigated with Monte Carlo methods
\cite {oka97,luo98a}.
Critical initial increase of the magnetization
is observed and the critical exponent $\theta$
is determined. However, as it is pointed out
in the cases of the Ising model, the Potts model
and the clock model \cite {sch96,cze96,oka97a},
 the determination of the static exponent
$\eta$ and the dynamic exponent $z$ from this process 
is not the best choice.
Sometimes the measurements of the auto-correlation
and/or the second moment is not so easy.
In comparison to this, the dynamic process
starting from {\it an ordered state} shows
its advantages for the measurements of the critical exponents
\cite {sch96,sta96a,sta92}.

In this letter, we numerically study the critical relaxation
of the two dimensional XY and 
FFXY model starting from {\it an ordered state}
and meanwhile determine the dynamic
exponent $z$ and the static exponent $\eta$
from the time-dependent Binder cumulant and magnetization.
Our results confirm the short-time dynamics scaling
and show the dynamic measurement of the critical exponents is
competitive with the traditional methods in equilibrium.

\begin{figure}[p]\centering
\epsfysize=12cm
\epsfclipoff
\fboxsep=0pt
\setlength{\unitlength}{1cm}
\begin{picture}(13.6,12)(0,0)
\put(0,0){{\epsffile{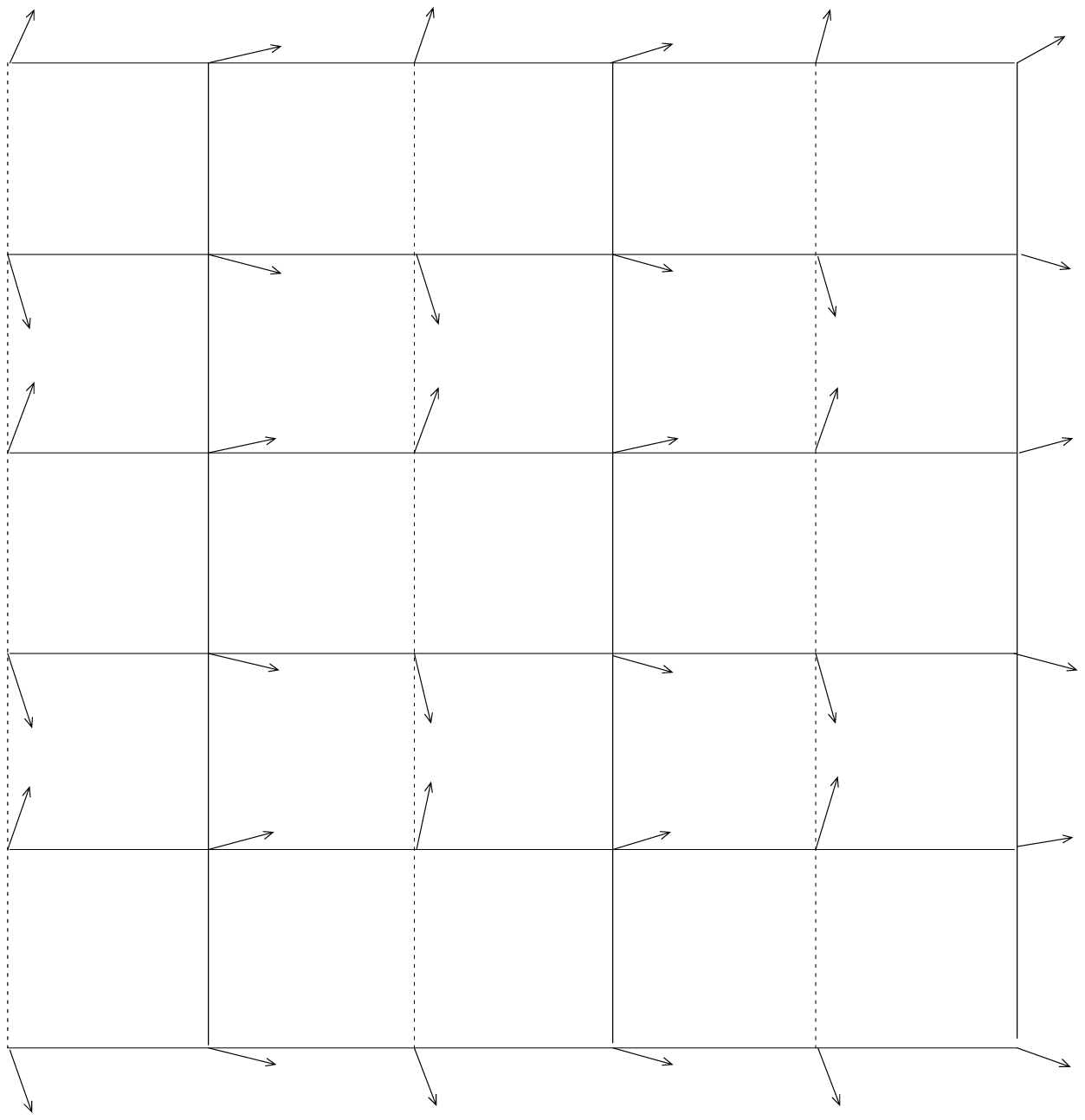}}}
\end{picture}
\caption{ A simple realization of the FFXY model and 
one of the ground states. The dotted vertical lines
represent the negative links with $f_{ij}=-1$.
}
\label{f1}
\end{figure}

The XY model and FFXY model in two dimensions 
can be defined by the Hamiltonian
\begin{equation}
H=K  \sum_{<ij>} f_{ij}\  \vec S_i \cdot \vec S_j\ ,
\label{e10}
\end{equation}
where $\vec S_i = (S_{i,x},S_{i,y})$ is a planar unit vector at site
$i$ and
the sum is over the nearest neighbours.
 In our notation the inverse temperature
 has been absorbed in the coupling $K$.
Here $f_{ij}$ take
the values $+1$ or $-1$, depending on the models.
For the XY model, $f_{ij} = 1$ on all links.
A simple realization of the FFXY model is by taking
$f_{ij}=-1$ on half of the vertical links (negative links)
and others are $+1$ (positive links).
This is shown in Fig.~\ref {f1}.
 The links maked by dotted lines represent
the negative links.

Let us first concentrate our attention on
the XY model. It is known that the XY model 
undergoes a Kosterlitz-Thouless phase
transition at a certain critical
temperature \cite {kos73,kos74}.
This critical temperature is numerically known
to be around  $T_c=1/K_c=0.90$ \cite {gup92}.
Below the critical temperature, no
real long range order emerges. The 
system remains critical
in the sense that the spatial correlation length
 is divergent, therefore, an universal dynamic scaling
form is expected in the whole time regime
below the critical temperature. For example,
for the $k$--th moment of the magnetization
the scaling form is 
\begin{equation}
M^{(k)}(t)=b^{-k\eta/2}
M^{(k)}(b^{-z}\ t).
\label{e20}
\end{equation} 
Taking $b=t^{1/z}$, one immediately
obtain the power law behaviour for the magnetization
$M(t) \equiv M^{(1)}(t)$
\begin{equation}
M(t) \sim t^{-\eta/2z}.
\label{e30}
\end{equation} 
Here $M(t)$ can be any components of the vector magnetization
defined as 
\begin{equation}
\vec M(t) = \frac {1}{L^2} \sum_{i} \vec S_{i}
\label{e40}
\end{equation}
with $L$ being the lattice size.

To determine $z$ {\it independently}, we introduce a {\it time-dependent}
 Binder cumulant 
$U(t,L) = M^{(2)}/M^2 -1$. Simple finite size
scaling analysis shows
\begin{equation}
U(t,L) \sim t^{d/z}.
\label{e50}
\end{equation}

To simulate the dynamic relaxation,
we take the initial state to be an ordered state,
i.e. $\vec S_i = (S_{i,x},S_{i,y}) = (1,0)$ for all
 spins. Then the system is suddently released to a dynamic evolution
of model A \cite {hoh77}
at the critical temperature or below.
 In this paper the Metropolis algorithm
is mainly adopted, while some simulations are also carried out
with the heat-bath algorithm to confirm universality.
We stop update at the Monte Carlo time step $t=750$
and then repeat the procedure. Average is taken over 
different random numbers. The total samples are $6\ 000$
for the lattice size $L=64$ and $32$, while $4\ 500$
for $L=128$. Errors are estimated by dividing the 
samples into three groups.

\begin{figure}[p]\centering
\epsfysize=12cm
\epsfclipoff
\fboxsep=0pt
\setlength{\unitlength}{1cm}
\begin{picture}(13.6,12)(0,0)
\put(1.2,8.0){\makebox(0,0){$M(t)$}}
\put(10.8,1.2){\makebox(0,0){$t$}}
\put(10.,7.5){\makebox(0,0){\footnotesize$Metropolis$}}
\put(8.5,4.2){\makebox(0,0){\footnotesize$heat-bath$}}
\put(1.8,2.0){\makebox(0,0){0.6}}
\put(0,0){{\epsffile{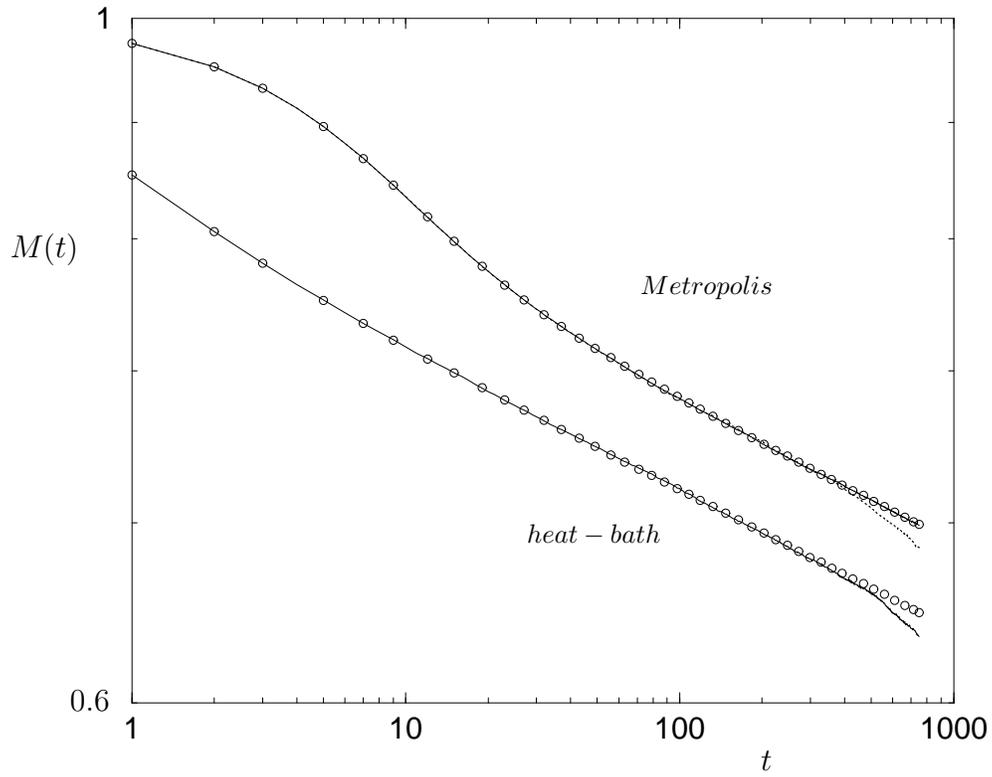}}}
\end{picture}
\caption{ The time evolution of the magnetization
for different lattice sizes and algorithms
is plotted in double-log scale for the XY model.
 The upper branch of the curves
is for the Metropolis algorithm while the lower one is
for the heat-bath algorithm. The dotted lines, solid lines
and circles correspond to the lattice size $L=32$, $64$
and $128$.
}
\label{f2}
\end{figure}

In Fig.~\ref{f2}, the time evolution
of the magnetization at the critical temperature
$T_c=1/K_c=0.90$ \cite {gup92}
with both the Metropolis and the heat-bath algorithm
is displayed for different lattice sizes.
$M(t)$ is the $x$ component of the magnetization $\vec M(t)$.
The $y$ component of the magnetization $\vec M(t)$ remains zero
since the initial value is zero.
The upper branch of the curves is the magnetization for the Metropolis
algorithm while the lower one is that for the heat-bath
algorithm. Dotted lines, solid lines and circles 
correspond to the lattice size $L=32$, $64$ and $128$.
At the very beginning of the time evolution,
the curves for the Metrpolis and the heat-bath algorithm
show quite different behaviour. After a certain time period,
however, universal power law behaviour indeed emerges
for both algorithms when the lattice size is 
 sufficiently big. For the Metrpolis algorithm,
the curves for $L=64$ and $128$ overlap completely
up to the time $t=750$.
The finite size effect for $L=64$ is already negligibly
small. In contrast to this, for the heat-bath algorithm
at around $t=300$ - $400$ the curve for $L=64$ starts
bending down and shows significant finite size effect.
This fact implies that the time scale of the heat-bath algorithm
is large than that of the Metropolis algorithm.
In other words, the time evolution with the heat-bath algorithm
is faster. This can also be seen from the microscopic time scale
$t_{mic}$, after which universal behaviour appears.
For the heat-bath algorithm $t_{mic} \sim 20$,
while for the Metropolis algorithm $t_{mic} \sim 80$.
Actually, the curve for the Metropolis algorithm with $L=32$
looks like that for the heat-bath algorithm
with $L=64$. Therefore the time scale for the heat-bath algorithm
is roughly a factor of four larger.

\begin{table}[h]\centering
$$
\begin{array}{|c|l|l|l|l|}
\hline
        &  \multicolumn{2}{c|} {Metropolis} &\multicolumn{2}{c|} {Heatbath}\\
\hline
  L  & \quad 64 &\quad 128 &\quad 64 &\quad 128\\
\hline
 \eta/2z & .0622(05) &  .0621(04) & .0637(13) & .0621(04)\\
\hline
\end{array}
$$
\caption{
 The exponent $\eta/2z$ measured 
with different lattice sizes and algorithms for the XY model.
}
\label{t1}
\end{table}

\begin{table}[h]\centering
$$
\begin{array}{|c|l|l|l|l|}
\hline
 T  & 0.90 & 0.86 & 0.80 & 0.70\\
\hline
 z & 1.96(4) & 1.98(4) & 1.94(2) & 1.98(4)\\
\hline
\eta & .244(5) & .212(4) & .178(2) & .143(3)\\
\hline
\end{array}
$$
\caption{ The exponent $z$ and $\eta$ measured for different temperatures
with the Metropolis algorithm for the XY model.
 The lattice size is $L=64$.}
\label{t2}
\end{table}

In the numerical simulations of the short-time dynamics,
is the heat-bath algorithm more efficient or 
the Metropolis algorithm? The advantage for the heat-bath
algorithm is that the microscopic time scale $t_{mic}$
is really small and there is clean universal scaling
behaviour in the very early time. However, the heat-bath
algorithm is in general more time comsuming. In Table~\ref {t1},
the measured exponent $\eta/2z$ for both algorithms
are listed. For the heat-bath algorithm,
the measurements are carried out in a time interval
$[50,300]$ while for the Metropolis algorithm
in a time interval $[200,750]$. In the numerical simulations
the update to $t=300$ for the heat-bath algprithm
and $t=750$ for the Metropolis algorithm need similar 
computer times. For the lattice size $L=128$,
both algorithms give the same results with the same quality.
However, for the lattice size $L=64$ the result
for the heat-bath algorithm is much worse.
The Metropolis algorithm shows its advantage in relatively
small lattice.

\begin{figure}[p]\centering
\epsfysize=12cm
\epsfclipoff
\fboxsep=0pt
\setlength{\unitlength}{1cm}
\begin{picture}(13.6,12)(0,0)
\put(1.2,9.5){\makebox(0,0){$U(t)$}}
\put(10.8,1.2){\makebox(0,0){$t$}}
\put(7.5,8.5){\makebox(0,0){\footnotesize$L=64$}}
\put(10.,5.7){\makebox(0,0){\footnotesize$L=128$}}
\put(0,0){{\epsffile{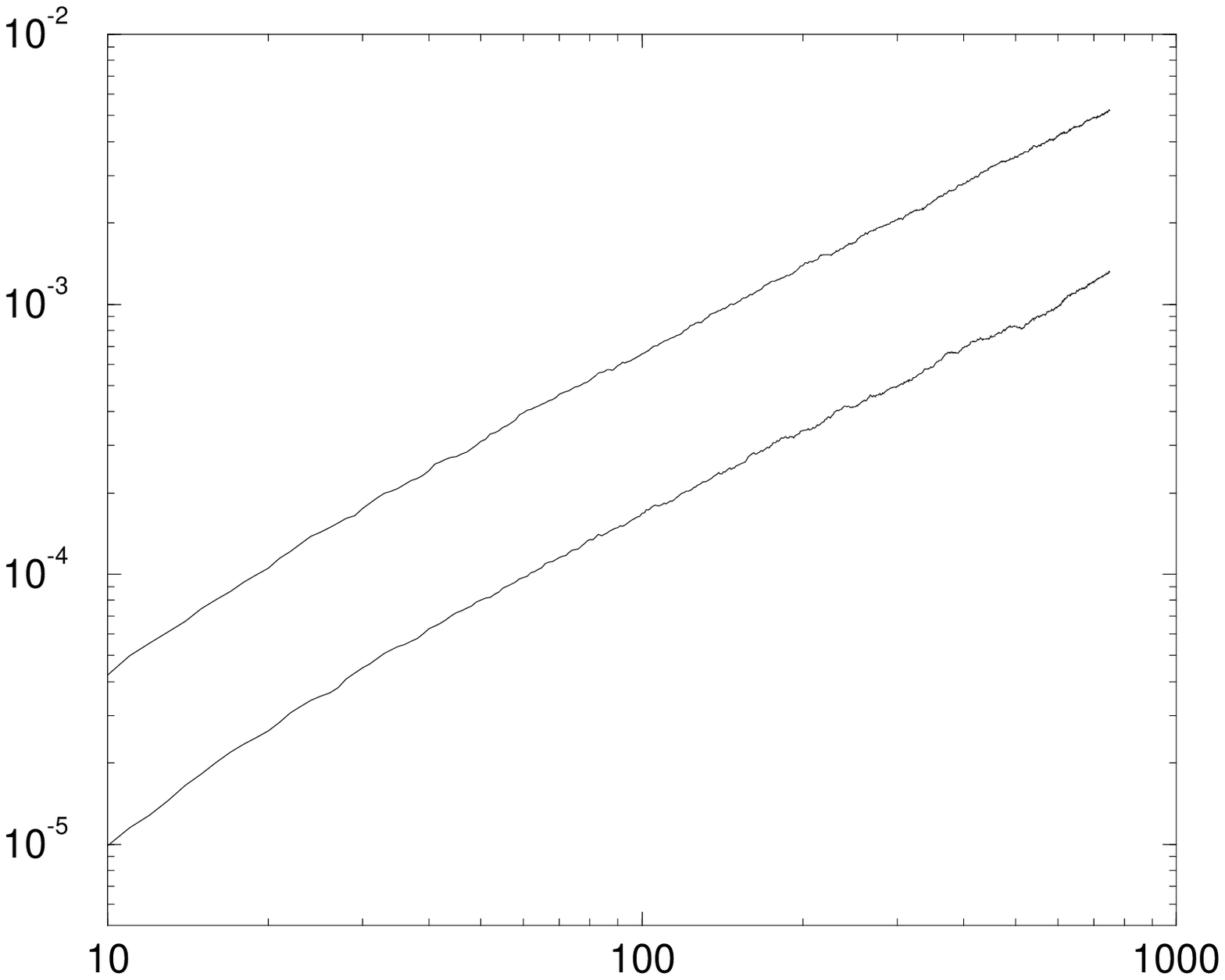}}}
\end{picture}
\caption{ The time-dependent Binder cumulant
for $L=64$ and $128$ with the Metropolis algorithm
is plotted in double-log scale for the XY model.}
\label{f3}
\end{figure}

From the power law decay of the magnetization, one can 
only obtain
the combination $\eta/2z$ of the exponents.
In order to determine the exponent $z$ independently,
we measure the time-dependent Binder cumulant $U(t,L)$. 
In Fig.~\ref{f3}, the cumulant $U(t,L)$
 at the critical temperature
with the Metropolis algorithm
is displayed for lattice size $L=64$ and $128$.
The measured exponent $z$ is $z=1.96(4)$ and $1.98(6)$
for the lattice size $L=64$ and $128$ respectively.
We see again that the finite size effect for $L=64$
is already negligibly small.
The simulations with the heat-bath algorithm
give similar results.

Compared with the dynamic process starting from
a disordered state, the determination of the critical exponents
from the dynamic relaxation from the ordered state
is more efficient. The lowest two of non-zero moments
in latter case
are $M^{(1)}$ and $M^{(2)}$ in contrast to
$M^{(2)}$ and $M^{(4)}$ in the former case. 
The values of $M^{(1)}(t)$ up to $t=750$
shown in Fig.~\ref {f2}
is still rather big and therefore the fluctuations are small.
The measurement of the cumulant
constructed from $M^{(1)}$ and $M^{(2)}$
is also easier than that constructed from
$M^{(2)}$ and $M^{(4)}$ in the former case.

\begin{figure}[p]\centering
\epsfysize=12cm
\epsfclipoff
\fboxsep=0pt
\setlength{\unitlength}{1cm}
\begin{picture}(13.6,12)(0,0)
\put(1.2,8.0){\makebox(0,0){$M(t)$}}
\put(10.8,1.2){\makebox(0,0){$t$}}
\put(10.,8.){\makebox(0,0){\footnotesize$T=0.70$}}
\put(8.5,3.7){\makebox(0,0){\footnotesize$T=0.90$}}
\put(1.8,2.0){\makebox(0,0){0.6}}
\put(0,0){{\epsffile{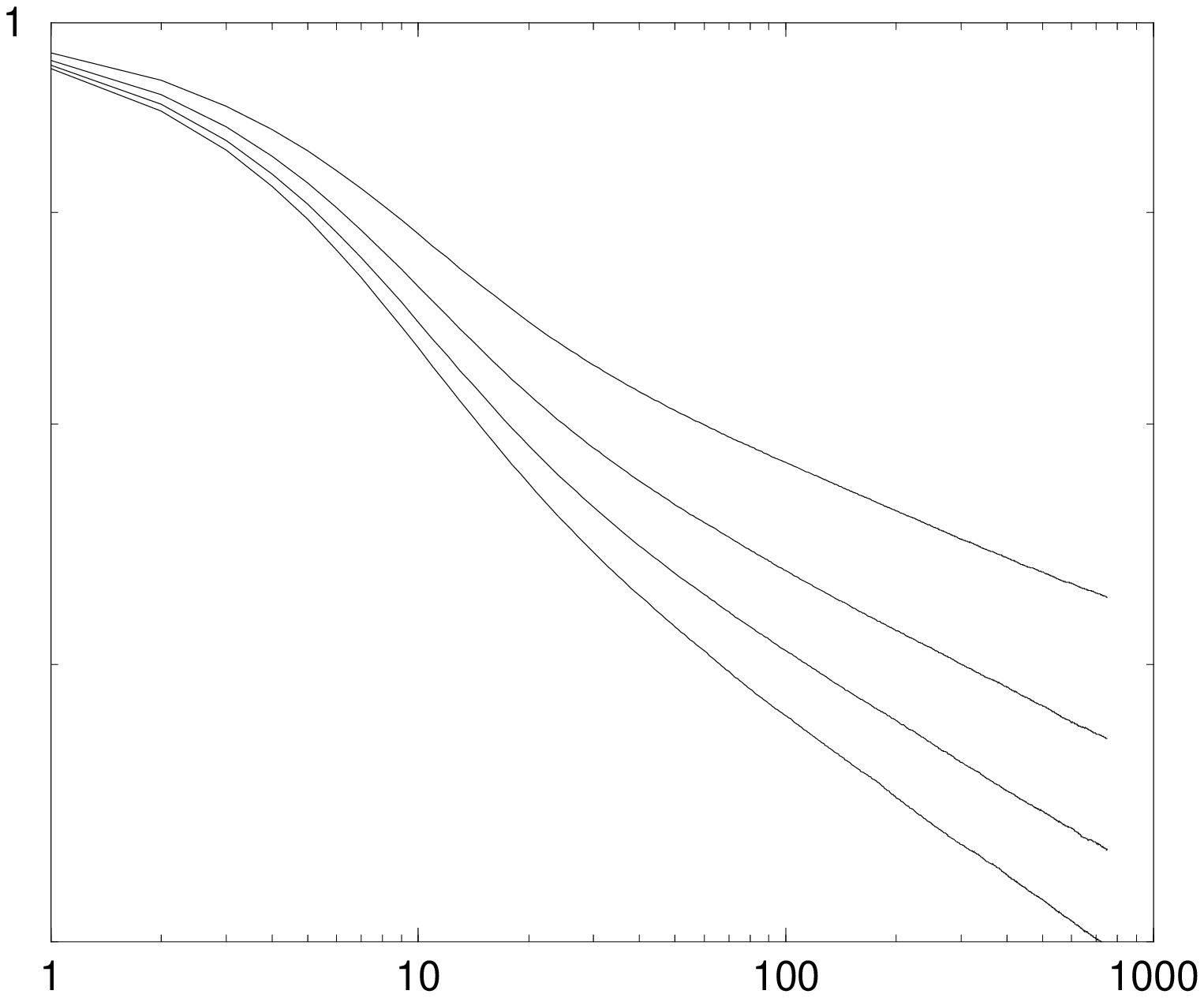}}}
\end{picture}
\caption{ The time evolution of the magnetization
for $L=64$ with different temperatures
is plotted in double-log scale for the XY model.
The Metropolis algorithm is used
in the simulations.
From above, the temperature
is $T=0.70$, $0.80$, $0.86$ and $0.90$ respectively.
}
\label{f4}
\end{figure}

Below the critical temperature, the XY model remains critical
and similar scaling is expected.
In Fig.~\ref{f4}, the time evolution
of the magnetization at different temperatures
with the Metropolis algorithm
is displayed for lattice size $L=64$.
All values for the exponent $z$ and $\eta$
measured with the Metropolis algorithm and
lattice size $L=64$ are given in Table~\ref {t2}.
With respect to
the temperature the exponent $z$ is rather clearly a constant 
very near $z=2$. This is consistent with the theoretical
prediction for the growth law in phase ordering
\cite {rut95}.
For the static exponent $\eta$, the results are
in very good agreement with those measured
in equilibrium. In Ref \cite {gup92},
for example, large scale Monte Carlo simulations
have been performed and it was reported that
 the static
exponent $\eta=0.238$ and $0.146$
for the temperature $T=0.90$ and $0.70$ respectively.
The agreement to the values in Table~\ref {t2}
is remarkable but our values of $\eta$ show
slightly more decreasing as the temperature decreases.
 The effort in our dynamic measurements
is really not much. Especially we need not
very big lattice in the numerical simulations
of the short-time dynamics. With the statistics of $6\ 000$
for a updating time $t=300$ or $t=750$,
the errors are already at least comparable with those
given in \cite {gup92}. Besides the static exponent
$\eta$, we also obtain the dynamic exponent $z$,
the measurement of which is very difficult
with traditional methods. Further, if the static exponent
$\eta$ is known, we can get even more rigorous
values for the dynamic exponent $z$ from the power law decay
of the magnetization $M(t)$ than that from the cumulant
given in Table~\ref {t2}.

\begin{table}[h]\centering
$$
\begin{array}{|c|l|l|l|l|l|}
\hline
 T  & 0.446 & 0.440 & 0.400 & 0.350 & 0.300\\
\hline
 z & 1.89(5) & 1.93(4) & 1.95(3) & 1.99(7) & 1.97(3) \\
\hline
\eta & .275(8) & .243(4) & .140(2) & .107(1) & .086(2) \\
\hline
\end{array}
$$
\caption{ The exponent $z$ and $\eta$ measured for different temperatures
with the Metropolis algorithm for the FFXY model.
 The lattice size is $L=64$.}
\label{t3}
\end{table}

\begin{figure}[p]\centering
\epsfysize=12cm
\epsfclipoff
\fboxsep=0pt
\setlength{\unitlength}{1cm}
\begin{picture}(13.6,12)(0,0)
\put(1.2,9.5){\makebox(0,0){$U(t)$}}
\put(10.8,1.2){\makebox(0,0){$t$}}
\put(8.5,8.5){\makebox(0,0){\footnotesize$T=0.446$}}
\put(8.5,4.3){\makebox(0,0){\footnotesize$T=0.300$}}
\put(0,0){{\epsffile{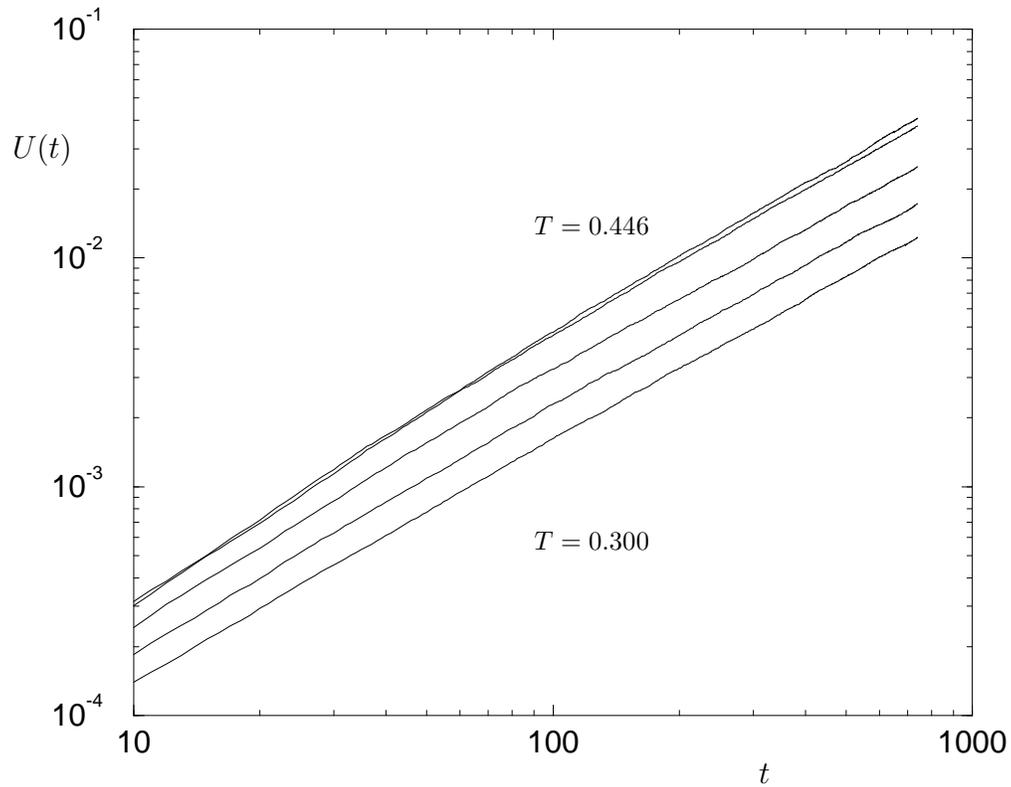}}}
\end{picture}
\caption{ The time-dependent Binder cumulant
for $L=64$ with different temperatures
is plotted in double-log scale for the FFXY model.
The Metropolis algorithm is used
in the simulations.
From above, the temperature
is $T=0.446$, $0.440$, $0.400$ , $0.350$ and $0.300$ respectively.
}
\label{f5}
\end{figure}

Encouraged by the success in the XY model, we have
also performed similar simulations for the FFXY model.
In equilibrium, due to the frustration of the couplings
the numerical simulation of FFXY model
is even more difficult than that for the XY model.
The Kosteritz-Thouless critical temperature $T_c$
is reported to be $T_c=0.446$ \cite {ram92,ram94,ols95}
or $T_c=0.440$ \cite {lee94}.
In the numerical simulations of the short-time dynamics,
 we take the initial ordered state
as the ground state shown in Fig.~\ref {f1}.
We measure the projection of the vector magnetization
$\vec M(t)$ in the initial direction and denote it 
also as $M(t)$. From $M(t)$ and the second moment of $\vec M(t)$
we construct the Binder cumulant.
In Fig.~\ref {f5}, the time-dependent Binder cumulant
for the lattice size $L=64$ with different temperatures
is plotted in double-log scale. It is clear that
nice power law behaviour appears for all the temperatures
after a microscopic time scale $t_{mic} \sim 100$.
The figure for the magnetization $M(t)$ looks similar as 
Fig.~\ref {f4}. All the measured exponents $z$ and $\eta$
for the lattice size $L=64$ and with
different temperatures are given in Table~\ref {t3}.
The statistics is the same as for the XY model.
From the errors in Table~\ref {t2} and \ref {t3},
we do not feel the simulations for the FFXY model
is more difficult than those for the XY model.
In Table~\ref {t3}, we see that the values for
the dynamic exponent $z$ are consistent with those
for the XY model. 
For the static exponent $\eta$, our values are slightly
bigger than those reported in Table V 
of Ref. \cite {ram94}. However, the finite size
effect of the results in Ref. \cite {ram94} is still
rather visible. For example, at the temperature
$T=0.440$ from the lattice size
$L=64$ to $96$ the value of $\eta$ changes from
$\eta=0.1666$ to $0.1955$. One easily expects
that for larger lattices the final value of $\eta$
may reach ours $\eta=0.243(4)$. In contrast to this,
the finite size
effect of our results is already negligible small. 

Finally we would like to mention that
our measured value for the dynamic exponent $z$
at the temperature $T=0.446$ is $z=1.89(5)$.
This is somehow smaller than that for other
temperatures below $T=0.446$. Further,
the measured value $\eta=0.275(8)$ for the exponent $\eta$
at this temperature is also apparently above the 
theoretical prediction $\eta=0.25$ at the 
critical temperature $T_c$. Therefore
our results suggest that the critical temperature
is near $T=0.440$ rather than $0.446$.
The measured value $\eta=0.243(4)$ at 
$T=0.440$ is very close to the value $\eta=0.244(5)$
for the XY model at the measured critical
temperature $T_c=0.90$ and the
theoretical prediction $\eta=0.25$
for the Kosterlitz-Thouless phase transition.

In conclusions, we have numerically investigated the universal
short-time behaviour of the critical relaxation
of the XY model and fully frustrated XY model
 starting from an ordered state.
We show that
even for more complex systems as the XY model and
 the FFXY model, we can obtain rigorously 
the dynamic exponent $z$ and the static exponent
from the short-time dynamics. Our measurements are
not hampered by critical slowing down.

{\it Acknowledgement:}
One of the authors (H.J. Luo) would like to thank the
Heinrich Hertz Stiftung for a fellowship.

\end{document}